\documentstyle[aps,12pt,preprint]{revtex}
\draft

\title{Relativistic Spheres
\thanks{The project supported by  National Natural 
Science Foundation of China. 1:lkhao@ieee.org, 2:frankwei@263.net, 3:liusiming@263.net.} }
\author{HAO Likun$^1$, WEI Jikun$^2$ and LIU Siming$^3$  \\  
{\small Department of Physics, Peking University,
Beijing 100871, China.}}
\date{\today}

\begin{document}
\maketitle

\begin{abstract}

By analyzing the Einstein's equations for the static 
sphere, we find that there exists a non-singular static configuration whose radius can approach its corresponding horizon size arbitrarily.

\end{abstract} 

If the energy density of a static perfect fluid (uncharged) sphere is assumed not to increase outwards, its radius $R$ must be larger than $9/8$ of its horizon size $R_g=2M$, where $M$ is its total mass in natural units, to avoid forming pressure singularity inside$^{\cite{Buch}}$. 
But a static sphere with its radius approaching its corresponding horizon size is of great interest not only for theoretical consideration$^{\cite{LynB}}$ but also to judge about the property of the matter in this extreme condition, namely close to turn into a black hole. 
In 1993, de Felice and Yu$^{\cite{Felice2}}$ found out one kind of such sphere which holds an inner singular boundary.
Although some meaningful analyses have been made based on that model$^{\cite{Felice3}\cite{Felice4}}$, the singular configuration is somewhat  exotic.
  
For the charged sphere, the electric force will weaken the pressure gradient and thus the pressure singularity may be avoided. 
In 1995, de Felice, Yu and Fang$^{\cite{Felice}}$ did find out a series of regular static configurations of the charged perfect fluid spheres whose radii $R$ can approach the external horizon sizes $R_+=M+\sqrt{M^2-Q_0^2}$ as $Q_0\to M$, where $Q_0$ is the charge of the spheres. 
However, we have proved recently that, for any given $Q_0/M<1$, the radius of the non-singular static configuration can not approach its external horizon size $R_+$ on the assumption that the sphere is made of perfect fluid$^{\cite{Yu}}$. 

It is well known that the transverse pressure of a solid shell can balance with its own gravity without the radial pressure gradient. 
Keeping this in mind, we analyze the Oppenheimer-Volkov (OV) equation for the static sphere and find that it is possible to construct the non-singular static configuration whose radius $R$ can approach $R_+$ for any assigned $Q_0/M<1$.

In what follows we shall use the natural units system with $c=G=1$ where $c$ is 
the velocity of light and $G$ the gravitational constant. Then in the spherically symmetric case, the metric takes the form
\begin{equation}
ds^{2}=-e^{\eta}dt^{2}+e^{\lambda}dr^{2}+r^{2}d\theta^{2}
+r^{2}sin^{2}\theta d\varphi^{2}. \label{g1}
\end{equation}

Only the regular static configuration will be considered, so the metric elements $e^{\lambda}$, $e^{\eta}$ are positive value functions of the coordinate radius $r$ only. And the Einstein's equations take on the form:
\begin{eqnarray}
-e^{-\lambda}(\frac{\lambda'}{r}-\frac{1}{r^{2}})-\frac{1}{r^{2}}
&=& 8\pi T^{t}_{t},  \label{g2} \\
e^{-\lambda}(\frac{\eta'}{r}+\frac{1}{r^{2}})-\frac{1}{r^{2}}
&=& 8\pi T^{r}_{r}, \label{g3} \\
\frac{e^{-\lambda}}{2}(\eta''+\frac{\eta'^{2}}{2}+\frac{\eta'-\lambda'}{r}
-\frac{\eta'\lambda'}{2})
&=& 8\pi T^{\theta}_{\theta}=8\pi T^{\varphi}_{\varphi}, \label{g5}
\end{eqnarray}
where ${T^{\nu}}_{\mu}$ are the energy-momentum tensor of the sphere, the 
primes here stand for the derivatives with respect to $r$.

Once we accept the Einstein's equations (\ref{g2})---(\ref{g5}), we 
see that the most general form of the energy-momentum tensor for a static sphere is:
\begin{equation}
{T^{\nu}}_{\mu}=diag\left[-\rho(r),\ p(r),\ F(r),\ F(r)\right].
\end{equation} 

We first discuss the uncharged sphere for which $\rho(r)$, $p(r)$ and $F(r)$ are its energy density, radial pressure and transverse pressure respectively. The equation of static equilibrium reads
\begin{equation}
p(r)'=-(\rho+p)\frac{\eta'}{2}+\frac{2}{r}(F-p). \label{g7}
\end{equation}
If we define a new variable $m(r)$ as
\begin{equation}
m(r)=\int^{r}_{0} 4\pi x^{2}\rho(x) dx, \label{g8}
\end{equation}
equation (\ref{g2}) can be solved as 
\begin{equation}
e^{-\lambda(r)}=1-\frac{2\ m}{r}, \label{g9}
\end{equation}
which suits the Schwarzschild metric at the boundary $r=R$. So $M=m(R)$ is the total mass of the sphere. 
Taking into account (\ref{g3}) and (\ref{g9}), (\ref{g7}) leads to the OV equation:
\begin{equation}
p_{(r)}'=-(\rho+p)\frac{(4\pi r p+\frac{m}{r^{2}})}{1-\frac{2m}{r}}+\frac{2}{r}(F-p). \label{g10}
\end{equation}

Because of the Bianchi identity, (\ref{g10}) can be obtained from the
Einstein's equations (\ref{g2})---(\ref{g5}). So usually in order to find the structure of the sphere and 
its internal metric, we need to know two equations of state: $p=p(\rho)$ and $F=F(\rho)$. But here,  we just want to show that it is possible to construct regular static configurations with $R\to R_g$. In the following, we shall consider the energy density $\rho(r)$ as an input.
For the regular static configuration, $e^{-\lambda(r)}$ must be positive, so the energy density $\rho(r)$ must satisfy
\begin{equation}
\int^{r}_{0} 4\pi x^{2} \rho(x) dx<\frac{r}{2}. \label{g11}
\end{equation}

Just as expected, equation (\ref{g10}) shows that the radial pressure gradient can be weakened by increasing the transverse pressure $F(r)$, a property we will use to construct the new models.

a: The most simple model of regular static configurations with $R$ approaching the horizon size $R_g=2M$ can be obtained by requiring the radial 
pressure $p(r)$ equal zero throughout the sphere. Since the radial pressure $p(r)$ should equal zero at the boundary, the requirement is a reasonable one. Then equation (\ref{g10}) shows that another equation of state must be 
\begin{equation}
F(r)=\frac{\rho\ m}{2r(1-\frac{2m}{r})}. \label{g12}
\end{equation}
We see, so long as $\rho(r)$ satisfies equation (\ref{g11}), there is no pressure singularity inside for any configurations with $R>R_g$.
 
In the Newtonian limit, we get
\begin{equation}
F(r)=\frac{\rho\ m}{2r}, \label{g13}
\end{equation}
which is the equilibrium equation of a static sphere with zero radial pressure in Newtonian theory. 

\medskip

\noindent b: In order to show the property quantificationally that the transverse pressure can weaken the radial pressure gradient, let us consider another specific model with $F(r)=\gamma p(r)$ and $\rho(r)=\rho_{0} $ where $\gamma$ and $\rho_{0}$ are two constant parameters. 
From equation (\ref{g10}), we see that $\gamma=1$ corresponds to the incompressible perfect fluid sphere and  the larger $\gamma$ is, the more slowly the radial pressure $p$ increase from the boundary inwards. 

Let $M=1$, equation (\ref{g8}) shows
\begin{equation}
\rho_0={3\over 4\pi R^3}. \label{g15}
\end{equation}
Then, for assigned $R$ and $\gamma$, one can solve equation (\ref{g10}) numerically to get the pressure inside the sphere. All these solutions can be represented by points in the ($R_g/R-\gamma$) plane. But not all of the solutions are physically acceptable. We define the regular static configurations as those which satisfy:

1: $R>R_g$,

2: There is no pressure singularity inside the sphere.

In Figure 1, we show the numerical results. Above the curve, there is pressure singularity inside the configuration, so the regular static configuration lies below the curve. We see ,with the increase of $\gamma$, the radii of the regular static configurations can approach the horizon size $R_g$ more closely. 
 
In fact, from equation (\ref{g10}), we have
\begin{equation}
F(r)=p+\frac{r}{2}[p'+\frac{(\rho+p)(4\pi rp+\frac{m}{r^{2}})}{1-\frac{2m}{r^{2}}}]. \label{g16}
\end{equation}
So, for arbitrarily assigned analytic functions of pressure $p(r)$ which satisfy the boundary condition $p(R)=0$ and energy density $\rho(r)$ which satisfy equation (\ref{g11}), there is an analytic function $F(r)$ which satisfy the OV equation (\ref{g10}). From equation (\ref{g3}), we have 
\begin{eqnarray}
e^{\eta(r)} &=& e^{\eta(R)}e^{-\int^{R}_{r}\frac{8\pi x^3p+2m}{x^{2} e^{-\lambda}}\ dx}. \label{g17}
\end{eqnarray}
We see that $e^{\eta(r)}$ is a positive function inside the sphere.

For the charged sphere, we just need to make a transformation$^{\cite{Beken}}:$
\begin{eqnarray}
\rho(r)&\longmapsto&\rho(r)+\frac{Q^{2}(r)}{8\pi r^{4}}, \label{g18}  \\
p(r)&\longmapsto&p(r)-\frac{Q^{2}(r)}{8\pi r^{4}}, \label{g19}  \\
F(r)&\longmapsto&F(r)+\frac{Q^{2}(r)}{8\pi r^{4}}, \label{g20}  \\
m(r)&\longmapsto&m(r)-{Q^2(r)\over 2r}, \label{g21}  \\
%e^{-\lambda}&\longmapsto&e^{-\lambda}-{Q^2(r)\over r^2}, \\
R_g&\longmapsto&R_+, 
\end{eqnarray}
 where $Q(r)$ is the electric charge within the radius $r$, $Q_0=Q(R)$ is the total charge of the sphere and $R_+=M+\sqrt{M^2-Q_0^2}$ is the external horizon size. One can show that all the analyses above are also valid except that the interior metric should suit the exterior Reissner-Nordstr\"om metric at the boundary.

Having regular static configurations with $R\to R_+$, we can show clearly why the total energy of a spherically symmetric static black hole solution is all outside the horizon size according to Lynden-Bell and Katz's (LK) definition of gravitational energy$^{\cite{LynB}}$. We have, in fact,
\begin{eqnarray}
E_m&=&\int_0^R e^{\eta(r)+\lambda(r)\over 2}4\pi r^2\left[\rho(r)+{Q^2(r)\over 8\pi r^4}\right] dr, \label{g22} \\
E_f&=&\int_0^R e^{\eta(r) \over 2}{1\over 2}(1-e^{-\lambda(r) \over 2})(1-e^{\lambda(r)\over 2}+r\eta'(r))dr, \label{g23}
\end{eqnarray}
where $E_m$  and $E_f$ are respectively the matter and electric field energy and the gravitational energy inside the sphere according to LK. Here we see, for regular static configurations, $E_m$ and $E_f$ will approach zero as $R\to R_+$, because equation (\ref{g17}) shows that the factor $e^{\eta(r)\over 2}$ in the integrals above approaches zero. So it is the red shift factor $e^{\eta(r)\over 2}$ that makes the energy inside the sphere tend to zero as $R\to R_+$.

From our models, it is obvious that the conjecture by Kristiansson$^{\cite{KSA}}$ is false, according to which the limit of regular embedding in the Euclidean space of conformally reduced space-like sections of the Reissner-Nordstr\"om metric coincides with the limit of regularity of the internal solutions.  

We conclude that there is a regular static configuration of the static sphere whose radius $R$ can approach the corresponding horizon size arbitrarily if one accepts that its radial pressure can be different from its transverse pressure. It is obvious that the sphere is not made of perfect fluid.

\newpage
\vskip 10mm
\centerline{\bf \large Figure Captions}
\vskip 1cm
\noindent
Fig.1. Configurations in the parameter space.
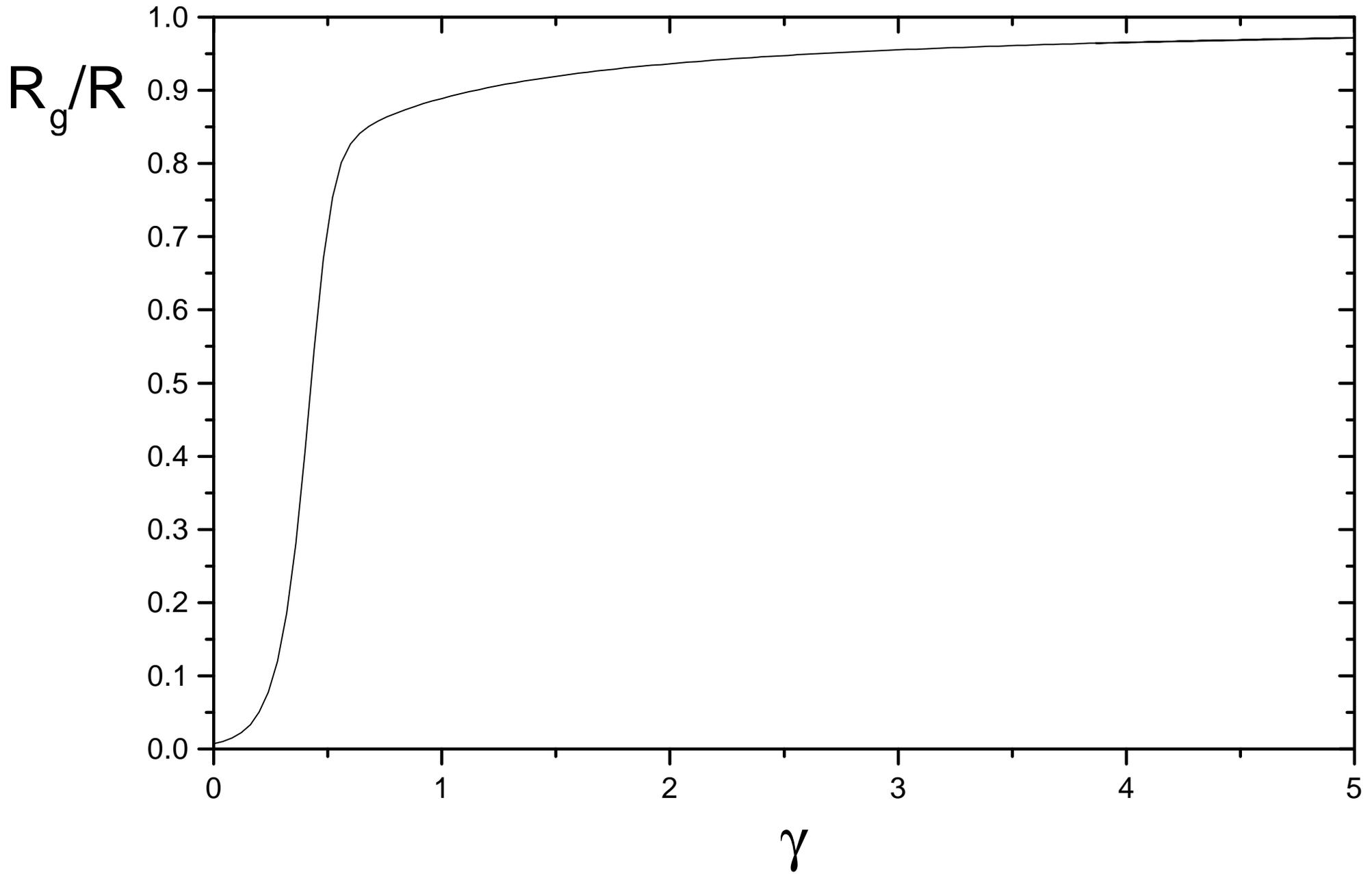
\end{document}